\documentclass[11pt]{article}
\usepackage{mathtools}
\usepackage[utf8]{inputenc}
\usepackage[T2A]{fontenc}
\usepackage[ukrainian,english]{babel}
\usepackage{indentfirst}
\usepackage{color}
\definecolor{gray50}{gray}{0.75}
\definecolor{gray25}{gray}{0.5}

\title{\textbf{Investigation of carbon dioxide phase shift possibility under extreme Antarctic winter conditions}}
\author{V.M.Vashchenko, 
		Ye.A. Loza \\ \\
		\texttt{State ecological academy}\\
		\texttt{of post graduate education and management}\\
		\texttt{Ukraine, Kyiv, V.Lypkivskogo str., 35}\\
		\texttt{Loza@bmyr.kiev.ua}}
\date{}
\begin{document}

\maketitle

\begin{abstract}
\noindent
The Antarctic winter atmosphere minimal temperature and pressure series reveal that $CO_2$ phase shift (deposition) is possible in some extreme cases, even leading to possible $CO_2$ snow phenomenon at Vostok Antarctic station and in other near South Pole regions. A hypothesis has been formulated that stable $CO_2$ snow cover might have formed in Earth past which may influence interpretation of glacial chronology records. This effect may also manifest in other minor gases. Its global climate role is discussed.

\end{abstract}

\section{Introduction}

There are two main gases in the Earth atmosphere - oxygen and nitrogen. However, minor gases, like $H_2O$ and $CO_2$ play a very important role in Earth global climate \cite{1}. The mole fractions and phase shifts temperatures of gases with mole fraction over 0.01\% are given in Table~1.

Oxygen and nitrogen have boiling and melting (freezing) points at temperatures of $-180$~$^oC$ and lower under normal atmospheric pressure. And so does argon.

However $H_2O$ experiences phase shifts in the atmosphere with well known consequences as rain and snow (starting from liquid or solid aerosol formation), having key influence on Earth global climate. Moreover the other aerosol particles may experience phase shifts under atmosphere temperature change \cite{2}. However, this phenomenon has not yet been well studied.

\begin{table}
\caption{Main atmospheric gases with mole fraction over 0.01\% \cite{3} and phase change temperatures at normal conditions.}
\begin{tabular}{ | c | c | c | c | c | p{3cm} |}
\hline
    \textbf{Gas} & \multicolumn{2}{|c|}{\textbf{Melting (sublimation)}} & \multicolumn{2}{|c|}{\textbf{Boiling}} & \textbf{Mole fraction}
\\
\hline
    $N_2$ & \textcolor{gray50}{-210~$^oC$} & \textcolor{gray50}{63~$^oK$} & \textcolor{gray50}{-196~$^oC$} & \textcolor{gray50}{77~$^oK$} & 78.08\%
\\
\hline
	$O_2$ & \textcolor{gray50}{-183~$^oC$} & \textcolor{gray50}{90~$^oK$} & \textcolor{gray50}{-219~$^oC$} & \textcolor{gray50}{54~$^oK$} & 20.95\%
\\
\hline
	$H_2O$ & 0~$^oC$ & 273~$^oK$ & +100~$^oC$ & 373~$^oK$ & $\approx$0.5\% average
\\
\hline
	Ar & \textcolor{gray50}{-189~$^oC$} & \textcolor{gray50}{84~$^oK$} & \textcolor{gray50}{-186~$^oC$} & \textcolor{gray50}{87~$^oK$} & 0.93\%
\\
\hline
	$CO_2$ & \textcolor{gray25}{-79~$^oC$} & \textcolor{gray25}{194~$^oK$} & n/a & n/a & 0.04\%
\\
\hline
	other & n/a & n/a & n/a & n/a & <0.001\% each
\\
\hline
\end{tabular}
\end{table}

This phenomenon may significantly influence the atmosphere state and the results of atmosphere investigations by optical methods and must be included in climate models.

Carbon dioxide cannot experience boiling/precipitation under any extreme atmospheric pressure values in natural conditions, requiring around 5200~mb with ever-observed natural maximum of 1100~mb.

In this paper we investigate $CO_2$ behavior at extreme cold temperatures. The registered lowest temperature record for the last century is $-89.2$~$^oC$ ("Vostok" station, Antarctic, 1983). As far as $CO_2$ sublimation/deposition temperature at 1000~mb pressure is $-78.5$~$^oC$ it is plausible that under such extreme conditions we may indeed witness $CO_2$ snow phenomenon at Earth South Pole.

\section{Atmosphere temperature ranges and other conditions}

First of all we should mention lowest temperature records close to $CO_2$ sublimation/deposition temperature found on Earth and summarized in table~2.

\begin{table}
\caption{Notable world low temperature records.}
\begin{tabular}{ | l | p{2.3cm} | p{2.1cm} | c | c |}
\hline
    \textbf{Location} & \parbox[t]{2cm}{\textbf{Lowest \\ temperature \\ record}} & \parbox[t]{3cm}{\textbf{Date of \\ the record}} & \parbox[t]{2cm}{\textbf{Site \\ height}} & \parbox[t]{2cm}{\textbf{Pressure\\ (appro-\\ximately)}}
\\
\hline
   South Pole & -82.8~$^oC$ & 23 Jun 1982 & 2800 m & ~ 800 mb
\\
\hline
	\parbox[t]{2cm}{Vostok \\ station \cite{4}} & \parbox[t]{2cm}{-89.2~$^oC$ \\ -91~$^oC$ \\ unconfirmed} & \parbox[t]{2cm}{21 Jul 1983 \\ winter 1997} & 3488 m & ~ 620 mb \cite{5}
\\
\hline
	\parbox[t]{2cm}{Oimekon \\ (Siberia)\cite{6}}  & \parbox[t]{2cm}{-67.7~$^oC$ \\ -76~$^oC$ \\ disputed} & 15 Jan 1885 & 800 m & ~ 900 mb
\\
\hline
\end{tabular}
\end{table}

We should also note that at different heights the atmosphere temperature is usually lower than that on the surface (e.g. by 45 to 75 degrees for height of ozone maximum which is 30-40~km). The pressure also reduces (e.g. from 1000 to 20-100~mb for the same height of ozone maximum). However, the real stratosphere zonal minimal temperatures at 70 mb heights measured by remote sounding is around $-91$~$^oC$, sometimes reaching $-93$~$^oC$ in the Antarctic \cite{7}.
We must also stress that all the atmospheric gases usually experience "supercooled" state. The temperatures required to form the snowflake may be by up to 40 degrees lower than the melting point.

The temperature anomalies may appear at some local points leading to appearance of some zones where $CO_2$ snow may form and then transport to areas with warmer conditions. The aerosol particles formed in the upper atmosphere under lower temperatures may quickly evaporate at the lower heights. I.e. we may be able observe quasi-stable snow forming and falling at a limited heights range.

Another important condition for aerosol and snow formation is availability of nucleation centers.

Concentrations of small, intermediate and large positive ions in Antarctica was estimated to be 2 to 6~$\times~10^2$~$cm^{-3}$, 7 to 30~$\times~10^2$~$cm^{-3}$, and from 5 to 12~$\times~10^3$~$cm^{-3}$ respectively \cite{8} at near-shore station Maitri which corresponds to average over-sea concentrations.

OH concentration in the Antarctic is higher than that at equator and constitutes about $2~\times~10^6$ molecules per cubic centimeter of air \cite{9}. Another favorable factor is that ionization radiation level at South Pole is higher due to Earth magnetic field properties.

Well-known $H_2O$ aerosols and snowflakes may also be nucleation centers for $CO_2$ deposition, and therefore increased quantity of $CO_2$ in regular snow is another possible manifestation of $CO_2$ phase shift under extremely cold temperatures. I.e. $CO_2$ aerosol forms and links to $H_2O$ snowflake, increasing overall $CO_2$ concentration in the snow layer and leading to ambiguity error in $CO_2$ glacial chronology records interpretation.

\section{$CO_2$ phase shift}

As far as this article is accented at Antarctic atmosphere, we should also mention more usual temperatures for "Vostok" station region. These are $-68.0$~$^oC$ average winter temperature and $-31.9$~$^oC$ average summer temperature. The average minimum temperature is $-71.6$~$^oC$ and average maximum temperature is $-50.3$~$^oC$ (with record at $-12.2$~$^oC$).

\begin{figure}
\begin{center}
\includegraphics[width=0.8\textwidth]{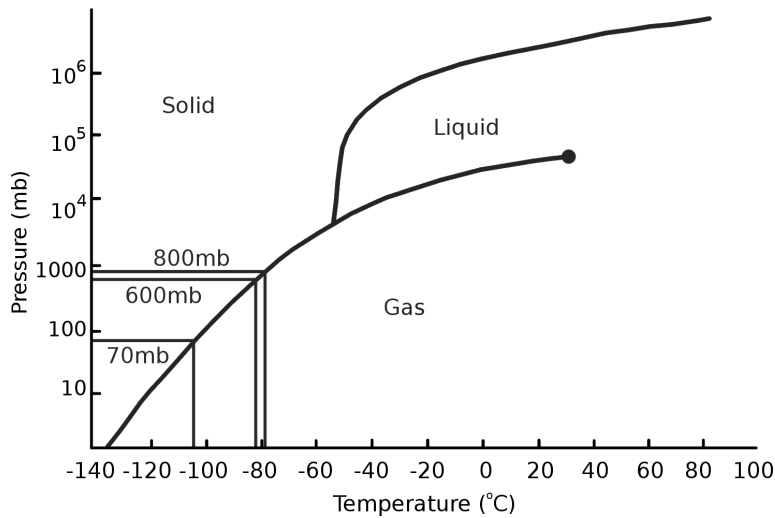}
\vskip-3mm\caption{$CO_2$ phase diagram, based on \cite{10}.}
\end{center}
\end{figure}

\begin{table}
\caption{$CO_2$ sublimation point pressure dependency based on fig.1}
\begin{tabular}{ | l | c |}
\hline
    \textbf{Pressure} & \textbf{Sublimation point temperature}
\\
\hline
    1000 mb (sea-level) & $-78.5$~$^oC$
\\
\hline
	800 mb (South Pole) & $-78.8$~$^oC$
\\
\hline
	600 mb (Vostok station) & $-82.4$~$^oC$
\\
\hline
	70 mb (Stratosphere) & $-104.8$~$^oC$
\\
\hline
\end{tabular}
\end{table}

As stated above, the stratosphere temperature at 70~mb height does not reach temperatures lower than $-93$~$^oC$ \cite{7}. Therefore, according to table~3 snow formation is impossible at these heights even at the record cold temperatures.

However, sublimation point temperature $-82.4$~$^oC$ at Vostok station, while exceeding average minimum temperature of $-71.6$~$^oC$, is quiet enough for extremal minimum temperatures, such as the 1983 temperature record of $-89.2$~$^oC$ with difference about 7 degrees.

Let's also note that South Pole with record minimal temperature of $-82.8$~$^oC$ and corresponding sublimation point temperature $-78.8$~$^oC$ is also a candidate for deposition of $CO_2$ with difference about 4 degrees.

Therefore, we may expect $CO_2$ phase shift as a rare event during extremely cold Antarctic winters.

\section{$CO_2$ snowflakes}

Another problem to consider is the detection possibility and characteristics of the $CO_2$ snowflakes. First of all, we should note that a regular $CO_2$ snowflakes size is estimated between 4 and 22 microns \cite{11}.

Moreover, all the theoretical and experimental research considering $CO_2$ snowflakes was made for Martian-like conditions (temperatures down to $-140$~$^oC$). By analogy with $H_2O$ snowflakes we may expect even smaller sizes of $CO_2$ snowflakes formed under conditions near to sublimation temperature. Therefore, the snowflake size should not exceed 5 microns. I.e. we should not witness regular snowflakes, but rather unseen by eye and even simple microscopic devices several microns size aerosol.

The globally averaged annual $H_2O$ precipitation is 990 millimeters \cite{15}. According to global average $CO_2$ concentration 0.04\% comparing to global average water vapor concentration around 1\% we may expect that $CO_2$ snow quantity generated during extreme cold period (let's consider it 24 hours) would hardly exceed 0.1 mm.

However, considering $CO_2$ snowflake size of 5 microns we may expect about 20 layers of snowflakes in the snow cover. In this case we should expect a tiny carbon dioxide rich layer of usual $H_2O$ snow.

Therefore $CO_2$ snowflakes may be definitely detected by: 1) electronic microscopy, leading to difficulty of "catching" and "transporting" the snowflakes; 2) by remote optical method such as spectropolarimetry \cite{12} leading to ambiguous results and 3) by nuclear-activation method on aerosol filters creating "transporting" and "storage" difficulty.

On the other hand super-cold chemical detection methods may be used e.g. detecting $CO_2$ gas level in sublimation camera after 10-30 microns filter, thou the average $CO_2$ snow quantity will be low, requiring high method sensitivity. Due to $CO_2$ gas density higher than that of average atmosphere density, the device filter should be placed at small angle (no greater than 45$^o$) with $CO_2$ accumulation \& sublimation camera located below the filter. Additional thermocouple is required to control the filter temperature not exceeding $CO_2$ sublimation temperature and sublimation camera temperature slightly exceeding this limit.

One more point to consider is that $CO_2$ snow is very likely to be mixed with regular $H_2O$ snow both in the air and in the snow cover. The pores in the upper snow layer about 0.1 mm wide will accumulate $CO_2$ snow quickly sublimating it due to higher internal temperature of the $H_2O$ snow layer.

On the other hand, in air $CO_2$ snow and $H_2O$ snow in the air are easily separated by simple aerosol filters because the majority of $CO_2$ snowflakes will be much smaller than most of the $H_2O$ snowflakes.

\section{Possibility of carbon dioxide stable snow cover at South Pole}

In case Earth global temperature (or just regional Antarctic temperature) drops by about 10 degrees, the $CO_2$ snow in Antarctic winter would appear on regular basis, followed by stable $CO_2$ winter snow in case of temperature reduction by 14 degrees.

As seen by EPICA and other glacial chronology records the Antarctica region show that Earth climate was close to these limits during some cold periods (see fig.~2).

Moreover temperature never actually reduces lower than 10 degrees than present average, as if there is a stable limit. This limit may be $CO_2$ stable snow cover formation. This effect may caused by neglecting stable $CO_2$ snow cover in ice records leading to significant temperature overestimate during such periods.

Different concentration of oxygen isotopes may be caused by admixture of $CO_2$ snow to regular $H_2O$ snow therefore causing misinterpretation of temperature in cryochronological records. And steady minimal temperature increase in the past may be caused by paleo $CO_2$ diffusing and leaving its original deposition layer and even diffusing to the surface and leaving snow cover with time.

This causes some degree of enrichment of regular snow with O and C isotopic compound different from normal $H_2O$ snow at the given temperature \cite{17,18}.

\begin{figure}
\begin{center}
\includegraphics[width=0.8\textwidth]{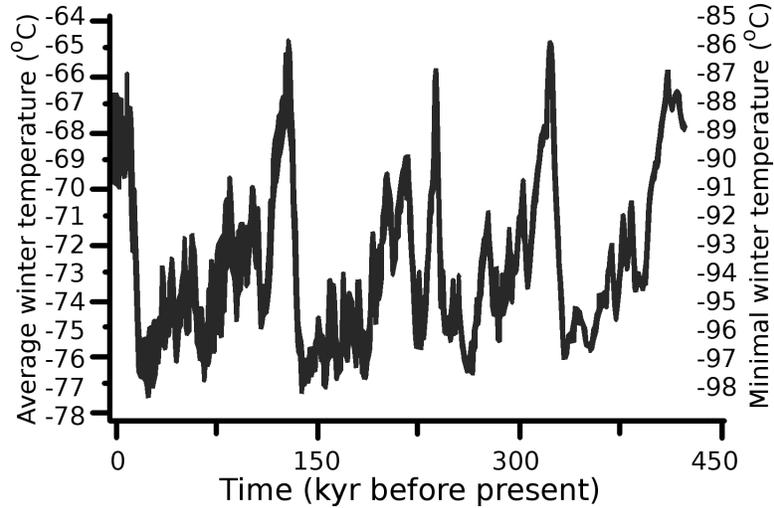}
\vskip-3mm\caption{Vostok Ice Core temperature records approximately adapted to current minimal winter temperatures based on \cite{16}.}
\end{center}
\end{figure}

\section{Minor gases}

In addition to $CO_2$ the following minor gases (with mole fraction over $10^{-7}~\%$) may experience phase shifts in the atmosphere during in-year temperature variations (Table~4 does not present a complete list).

\begin{table}
\caption{Minor gases molar concentrations, melting and boiling temperatures.}
\begin{tabular}{ | c | c | c | c | c | l |}
\hline
    \textbf{Gas} & \multicolumn{2}{|c|}{\textbf{Melting (sublimation)}} & \multicolumn{2}{|c|}{\textbf{Boiling}} & \textbf{Mole fraction}
\\
\hline
    $O_3$ & \textcolor{gray50}{-192~$^oC$} & \textcolor{gray50}{81~$^oK$} & \textcolor{gray25}{-112~$^oC$} & \textcolor{gray25}{161~$^oK$} & $10^{-5}~\%$
\\
\hline
	Xe & \textcolor{gray25}{-112~$^oC$} & \textcolor{gray25}{161~$^oK$} & \textcolor{gray25}{-108~$^oC$} & \textcolor{gray25}{165~$^oK$} & $10^{-5}~\%$
\\
\hline
	$N_2O$ & -91~$^oC$ & 182~$^oK$ & -88~$^oC$ & 185~$^oK$ & $10^{-4}~\%$
\\
\hline
	$NO_2$ & \textbf{-11~$^oC$} & \textbf{262~$^oK$} & \textbf{21~$^oC$} & \textbf{294~$^oK$} & $10^{-5}~\%$
\\
\hline
	$HF$ & -84~$^oC$ & 189~$^oK$ & \textbf{20~$^oC$} & \textbf{293~$^oK$} & less than $10^{-5}~\%$
\\
\hline
	$Hg$ & \textbf{-30~$^oC$} & \textbf{243~$^oK$} & \textcolor{gray50}{357~$^oC$} & \textcolor{gray50}{630~$^oK$} & less than $10^{-5}~\%$
\\
\hline
	$NH_3$ & -78~$^oC$ & 195~$^oK$ & \textbf{-33~$^oC$} & \textbf{240~$^oK$} & less than $10^{-5}~\%$
\\
\hline
\end{tabular}
\end{table}

While such gases as $O_3$ and Xe may hardly form phase shifts even under extreme temperature records, thou some tiny aerosol particles may form locally, $N_2O$ may experience a phase shift in the atmosphere at extreme conditions. Moreover, $HF$, $Hg$ and $NH_3$ have phase shift temperatures that stably appear in a large fraction of the Earth surface. And $NO_2$, like $H_2O$, may experience two phases shifts in a large fraction of the Earth surface.

Thou their concentration is far smaller than that of $CO_2$, such phase shifts definitely influence global climate by stabilizing respective temperatures at certain phase shifts temperature values and by changing atmosphere chemistry and radiative parameters. Moreover, formation of such quasi-seasonal atmospheric aerosol changes must be included in atmospheric models used for Earth atmosphere remote measurements.

However, investigation of these phase shift manifestation is outside of the scope of this article.

\section{Conclusions}

1. An overview of possibility of phase shifts of minor gases in the Earth atmosphere was made with accent at carbon dioxide. It was found that $CO_2$ may form solid aerosol and even microscopic snowflakes during extreme Antarctic winter conditions at South Pole and Vostok station.

2. A hypothesis has been formulated that stable $CO_2$ snow cover might have formed in Earth past which may influence interpretation of glacial chronology records.

3. A brief analysis of other minor gases phase shifts conditions has been made. It was found that $N_2O$, $HF$, $Hg$, $NH_3$ and $NO_2$ may also experience phase shifts in Earth atmosphere temperature range. Seasonal changes in atmosphere aerosol composition and aggregate state may impact the results of remote optical observations of the Antarctic atmosphere and should be included in climate models.


\begin{thebibliography}{18}
\bibitem{1}Русов В.Д., Глушков А.В., Ващенко В.Н. Астрофизическая модель глобального климата Земли // Киев:Наукова думка, 2003, 214 с.
\bibitem{2}C. L. Badger, I. George, P. T. Grifiths, C. F. Braban, R. A. Cox, and J. P. D. Abbatt Phase transitions and hygroscopic growth of aerosol particles containing humic acid and mixtures of humic acid and ammonium sulphate // Atmos. Chem. Phys., 6, 755–768, 2006.
\bibitem{3}Pidwirny, M. (2013). Atmospheric composition. Retrieved from http://www.eoearth.org/view/article/150296.
\bibitem{4}А. Б. Будрецкий. Новый абсолютный минимум температуры воздуха // Информационный Бюллетень Советской Антарктической Экспедиции № 105. Ленинград, Гидрометеоиздат, 1984 г.
\bibitem{5}Station Vostok // Russian Antarctic Expedition http://www.aari.aq/stations/vostok/vostok\_en.html.
\bibitem{6}Stepanova N. On the lowest temperatures on Earth // Monthly weather review, January, 1958, p.6-10.
\bibitem{7}Stratosphere global temperature time series // NOAA, National Weather Service http://www.cpc.ncep.noaa.gov/products/stratosphere/temperature/.
\bibitem{8}Devendraa Siingh, Vimlesh Pant, A K Kamra. Measurements of positive ions and air-earth current density at Maitri, Antarctica //  	arXiv:0905.4927 [physics.ao-ph].
\bibitem{9}J. Sanders. Scientists Find Evidence of Highly Oxidizing Environment Over the South Pole // July 2001, Georgia Institute of Technology Research, http://gtresearchnews.gatech.edu/newsrelease/SPOLE.html.
\bibitem{10}Good plant design and operation for onshore carbon capture installations and onshore pipelines - 2.1 General properties and uses of carbon dioxide // Energy Institute, London, September 2010 (first edition), 137p.
\bibitem{11}The Dry Ice 'Snowflakes' of Mars. Analysis by Ian O'Neill // Discovery news http://news.discovery.com/space/the-foggy-carbon-dioxide- snow-of-mars-120619.html.
\bibitem{12}О.В. Мороженко. Методи і результати дистанційного зондування планетних атмосфер // К. : Наук. думка, 2004. - 647 c.
\bibitem{15}Dr. Chowdhury's Guide to Planet Earth (2005). "The Water Cycle". WestEd. http://www.planetguide.net/book/chapter\_2/water\_cycle.html
\bibitem{16}Petit et. al. Vostok Ice core temperature records
\bibitem{17}Pamela A. Gemery, Michael Trolier, James W. C. White. Oxygen isotope exchange between carbon dioxide and water following atmospheric sampling using glass flasks // Journal of Geophysical Research: Atmospheres Vol. 101, Issue D9, pages 14415–14420, 20 June 1996.
\bibitem{18}Deniro MJ, Epstein S. Relationship between the oxygen isotope ratios of terrestrial plant cellulose, carbon dioxide, and water. Science. 1979 Apr 6;204(4388):51-3. PubMed PMID: 17816736.
\end{thebibliography}
\end{document}